\documentclass[11pt]{article}
\usepackage{amsmath}
\usepackage{epsfig,bm}
\begin{document}

\title{NON-INTEGER QUANTUM TRANSITIONS\thanks{ Project supported by
the National Nature Science Foundation of China (Grant
No.10305001).}}

\author{Qi-Ren Zhang
\\CCAST(World Lab),P.O.Box 8730,Beijing,100080\\and\\
Department of Technical Physics, Peking University ,
Beijing,100871,China\thanks{Mailing address.}}

\maketitle

\vskip0.3cm

\begin{abstract}
We show that in the quantum transition of a system induced by the
interaction with an intense laser of circular frequency $\omega$,
the energy difference between the initial and the final states of
the system is not necessarily being an integer multiple of the
quantum energy $\hbar\omega$.
\end{abstract}



\section{Introduction}
It is widely accepted that the Bohr condition
\begin{eqnarray}
E_2-E_1=\pm \hbar\omega \label{1}
\end{eqnarray}
expresses the energy conservation in the quantum transition from
the state with energy $E_1$ to the state with energy $E_2$,
induced by the interaction of the system with the electromagnetic
field of circular frequency $\omega$. It may be generalized to
\begin{eqnarray}
E_2-E_1=N\hbar\omega \; ,\label{2}
\end{eqnarray}
with an integer $N$, when the system is interacting with a laser.
This generalized Bohr condition is still thought to be an
expression of the energy conservation in the transition, with the
number of absorbed or emitted photons being more than 1. The
transitions satisfying (\ref{2}) with $N>1$ had been observed
experimentally in forms of the multi-photon ionization
(MPI)\cite{v} and the above threshold ionization (ATI)\cite{a} .
Here, we would emphasize that the Bohr condition (\ref{1}) or
(\ref{2}) is approximate, and its energy conservation
interpretation is not exact either. As we know, every spectrum
line has its width. It means there is always an error when
(\ref{1}) is applied to an individual transition. It becomes
specially obvious when we apply it to the magnetic resonances. In
this case, the resonance frequency is determined by a constant
magnetic field, and the width of resonance is determined by a
rotating magnetic field. The strengths of these two fields are
comparable. It means, in most individual magnetic transitions the
Bohr condition (\ref{1}) is seriously violated. However, the
violation of Bohr condition does not mean the violation of energy
conservation. Since the energy conservation means that the total
energy of the system and the electromagnetic fields does not
change, but the sum of the energy of the system and that of the
electromagnetic fields is not the total energy. Their difference
is the interaction energy between the system and the
electromagnetic fields. Only when this difference is negligible,
the Bohr condition becomes a good approximation of energy
conservation, and therefore has to be fulfilled. In this case the
interaction energy may be regarded as a perturbation. It is
realized for the transitions in weak fields. In the following, we
shall see, for the transitions in lasers, Bohr condition may be
badly violated. For an individual transition we always have
\begin{eqnarray}
E_2-E_1=\eta \hbar\omega \; ,\label{3}
\end{eqnarray}
with $\eta $ being defined in it. For transitions in strong
electromagnetic fields, like in lasers, $\eta$ may be quite
different from any integer. We call this kind of transition a
non-integer quantum transition.

Therefore, Bohr condition is not a first principle, but a special
relation for special processes. It may be deduced from quantum
mechanics by perturbation. (\ref{2}) is a result of the limit
\begin{eqnarray}
\lim_{t\rightarrow
\infty}\frac{\sin^2[(E_2-E_1-N\hbar\omega)\frac{t}{2\hbar}]}
{(E_2-E_1-N\hbar\omega)^2\frac{t}{2\hbar}}=\pi\delta(E_2-E_1-N\hbar\omega)
\label{4}
\end{eqnarray}
in the $N$th order perturbation,  showing that the Bohr condition
is a representation of the resonance with an integer $N$. In a
strong electromagnetic field, the interaction energy between the
system and the field is large. The quantum transition has to be
handled by non-perturbation method. This kind of resonance may not
appear and the non-integer quantum transition appears. It is a
non-perturbation effect.

\section{Transitions between discrete levels, laser Raman effects\label{s1} }
A laser is a classical limit of the intense electromagnetic wave.
In the Coulomb gauge, the circularly polarized laser is therefore
well described by the vector potential
\begin{eqnarray}
{\textit{\textbf{A}}}=A[{\textit{\textbf{x}}}_0\cos(kz-{\omega}t)
+{\textit{\textbf{y}}}_0\sin(kz-{\omega}t)]\  .\label{11}
\end{eqnarray}
Consider the quantum transition of a hydrogen atom irradiated by
this laser. At the moment, we would simplify the problem to the
motion of a non-relativistic spin-less electron in the Coulomb
field and the laser. Possible corrections of the omitted effects
on the result will be discussed in section \ref{s4}. The
Hamiltonian of this electron is
\begin{eqnarray}
\hat{H}=\hat{H}_0+\hat{H}^\prime \ ,\label{12}
\end{eqnarray}
with
\begin{eqnarray}
\hat{H}_0&=&\frac{\hat{p}^2}{2m}+V(r)\ ,\label{13}\\
\hat{H}^\prime &=&\frac{eA}{m} [\hat{p}_x\cos(kz-\omega
t)+\hat{p}_y\sin(kz-\omega t)]+\frac{e^2A^2}{2m} \ .\label{14}
\end{eqnarray}
$V(r)=-\alpha\hbar c/r$ is the Coulomb potential for the electron,
$\alpha $ is the fine structure constant, $-e$ and $m$ are the
electric charge and mass of the electron respectively. The
Schr\"odinger equation
\begin{eqnarray}
{\rm i}\hbar\frac{\partial\Psi}{\partial t}=\hat{H}\Psi \label{15}
\end{eqnarray}
for the electron is time dependent. However, a transformation
\begin{eqnarray}
\Psi({\textit{\textbf r}},t)={\rm e}^{{\rm i}\omega\hat{L}_zt
/\hbar}\Phi({\textit{\textbf r}},t)\label{16}
\end{eqnarray}
changes it into a time independent pseudo-Schr\"odinger equation
\begin{eqnarray}
{\rm i}\hbar\frac{\partial\Phi}{\partial t}=\hat{H}_{\rm ps}\Phi\
, \label{17}
\end{eqnarray}
with the pseudo-Hamiltonian
\begin{eqnarray}
\hat{H}_{\rm ps}=\hat{H}_0^\prime +\hat{H}^{\prime\prime}\
,\label{18}
\end{eqnarray}
in which
\begin{eqnarray}
\hat{H}_0^\prime=\hat{H}_0+\omega\hat{L}_z \;\;\;\;
\mbox{and}\;\;\;\; \hat{H}^{\prime\prime}=\frac{eA}{m}
[\hat{p}_x\cos(kz)+\hat{p}_y\sin(kz)]+\frac{e^2A^2}{2m} \label{20}
\end{eqnarray}
are time independent. Denote the $i$th eigenfunction of
$\hat{H}_{\rm ps}$ by $\phi_i({\textit{\textbf r}})$. We have
\begin{eqnarray}
\hat{H}_{\rm ps}\phi_i({\textit{\textbf
r}})=E_i\phi_i({\textit{\textbf r}})\ ,\label{21}
\end{eqnarray}
$E_i$ is the pseudo-energy of the electron in the
pseudo-stationary state $\phi_i({\textit{\textbf r}})$. They may
be quite different from the energy $E_n=-\alpha^2mc^2/2n^2$ and
the stationary state
\begin{eqnarray}
\psi_{nl\mu}({\textit{\textbf r}})=\frac{{\rm
i}^l}{(2l+1)!}\left[\left(\frac{1}{na_0}\right)^{2l+3}\frac{(n+l)!}{2n(n-l-1)!}\right]^{1/2}
{\rm e}^{-\frac{r}{na_0}}r^l{\rm
F}(l+1-n,2l+2,\frac{2r}{na_0}){\rm
Y}_{l\mu}(\theta\varphi)\label{22}
\end{eqnarray}
of the electron in an isolated hydrogen atom respectively. $a_0$
is the Bohr radius, F is the confluent hypergeometric function,
and Y is the spherical harmonic function. $r,\theta$ and $\varphi$
are spherical coordinates of the electron. Now, let us expand
$\phi_i$ in terms of the wave functions [$\psi_{nl\mu}$]:
\begin{eqnarray}
\phi_i({\textit{\textbf
r}})=\sum_{nl\mu}C_{nl\mu}(i)\psi_{nl\mu}({\textit{\textbf r}})\
.\label{23}
\end{eqnarray}
This is an approximation, since the set [$\psi_{nl\mu}$] of bound
states only is not complete. We expect that it is good enough for
the state $\phi_i$ near a low lying  bound state. We further
assume that in the expansion (\ref{23}) only terms with $n\le n_0$
are important, therefore one may truncate the summation on the
right at $n=n_0$. This makes the eigen-equation (\ref{21}) become
an linear algebraic equation, and therefore may be solved by the
standard method\cite{w}.

The factor ${\rm i}^l$ on the right of (\ref{22}) makes the matrix
elements of $\hat{H}_{\rm ps}$ be real in the $\hat{H}_0^\prime$
representation . Therefore the solutions $[C_{nl\mu}(i)]$ are
real. We have the reciprocal expansion
\begin{eqnarray}
\psi_{nl\mu}({\textit{\textbf
r}})=\sum_iC_{nl\mu}(i)\phi_i({\textit{\textbf r}})\ .\label{24}
\end{eqnarray}
Suppose the hydrogen atom stays in the state $\psi_{nl\mu}$ when
$t\leq 0$. The laser arrives at $t\!=\!0$. According to
(\ref{16}), (\ref{17}) and (\ref{21}), at $t>0$, the pseudo-state
will be
\begin{eqnarray}
\Phi({\textit{\textbf
r}},t)=\sum_iC_{nl\mu}(i)\phi_i({\textit{\textbf r}}){\rm
e}^{-{\rm
i}E_it/\hbar}=\sum_{n'l'\mu'}\sum_iC_{nl\mu}(i)C_{n'l'\mu'}(i){\rm
e}^{-{\rm i}E_it/\hbar}\psi_{n'l'\mu'}({\textit{\textbf r}})\ ,
\label{25}
\end{eqnarray}
and the state becomes
\begin{eqnarray}
\Psi({\textit{\textbf
r}},t)=\sum_{n'l'\mu'}\sum_iC_{nl\mu}(i)C_{n'l'\mu'}(i){\rm
e}^{-{\rm
i}(E_i-\mu'\hbar\omega)t/\hbar}\psi_{n'l'\mu'}({\textit{\textbf
r}})\ . \label{251}
\end{eqnarray}
The transition probability of the hydrogen atom from the state
$\psi_{nl\mu}$ to the state $\psi_{n'l'\mu'}$ is
\begin{eqnarray}
w_{n'l'\mu';nl\mu}(t)=\left| \sum_iC_{nl\mu}(i)C_{n'l'\mu'}(i){\rm
e}^{-{\rm i}(E_i-\mu'\hbar\omega)t/\hbar}\right|^2 \ . \label{26}
\end{eqnarray}
It is a multi-periodic function of $t$. The periods are of the
microscopic order of magnitude. On the other hand, the observation
is done in a macroscopic duration. Therefore the observed
transition probability is a time average of (\ref{26}) over its
periods. The averages of the cross terms with different $i$ in the
summation are zeros. It makes the observed transition probability
be
\begin{eqnarray}
W_{n'l'\mu':nl\mu}=\sum_iC_{nl\mu}^2(i)C_{n'l'\mu'}^2(i)\
.\label{27}
\end{eqnarray}
From the normalizations
\begin{eqnarray}
\sum_{nl\mu}C_{nl\mu}^2(i)=1\;\;\;\;\;\mbox{and}\;\;\;\;\;
\sum_iC_{nl\mu}^2(i)=1\label{28}
\end{eqnarray}
one sees the normalization
\begin{eqnarray}
\sum_{n'l'\mu'}W_{n'l'\mu';nl\mu}=1\ .
\end{eqnarray}
It shows that the expression (\ref{27}) for the transition
probability is reasonable.

When the amplitude $A$ is small (weak light), one may solve
equation (\ref{21}) by perturbation. The unperturbed
pseudo-Hamiltonian is $\hat {H}_0^\prime$, the unperturbed
pseudo-states are $[\psi_{nl\mu}]$, with unperturbed
pseudo-energies $[E_n+\mu\hbar\omega]$, and the perturbation is
$\hat {H}^{\prime\prime}$. In optic problems, wave length is
usually much longer than the Bohr radius, therefore $ka_0<< 1$.
Under these conditions, the perturbation becomes
\begin{eqnarray}
\hat{H}^{\prime\prime}=\frac{eA}{m}\hat{p}_x\ .\label{280}
\end{eqnarray}
The selection rules of its non-zero matrix elements include
\begin{eqnarray}
\Delta\mu\equiv\mu'-\mu=\pm 1\ .\label{281}
\end{eqnarray}
If the Bohr condition
\begin{eqnarray}
E_{n'}-E_n=\pm\hbar\omega \label{29}
\end{eqnarray}
is fulfilled, pseudo-states $\psi_{nl\mu}$ and $\psi_{n'l'\mu'}$
with $\Delta\mu=\pm 1$ are degenerate. The correct zeroth-order
approximation of eigen-states of $\hat{H}_{\rm ps}$ has to be
formed by their superpositions. The problem is equivalent to an
eigenvalue problem of a two level system. In the limit of
$A\rightarrow 0$, a resonance factor of type (\ref{4}) with
$N=\pm1$ appears. On the contrary, if the condition (\ref{29}) is
not fulfilled, the transition probability is zero in the
zeroth-order perturbation. We see, the transition probability
calculated by (\ref{27}) is  in agreement with that obtained by
the traditional method. This result may be regarded as a check of
the method proposed here. Now let us use it to consider the
transitions in lasers.

The $\hat{H}_0^\prime$ representation of $\hat{H}_{\rm ps}$, after
being truncated at $n_0=18$, is a $2109\times 2109$ matrix. It is
solved numerically by the standard method\cite{w} for various
values of $A$ and $k$. Substituting the solved eigen-vectors into
(\ref{27}), we obtain transition probabilities for these cases.
The results are shown in the following figures.

\begin{figure}[h]
\centerline{\epsfig{file=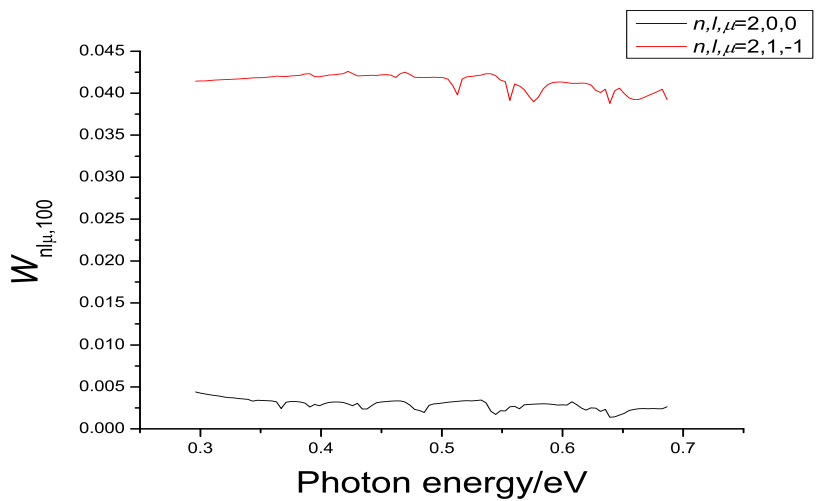,width=12cm, height=7.5cm}}
\caption{Transition probabilities of a hydrogen atom interacting
with a circularly polarized laser of
$A=5\times10^{-6}$V$\cdot$s/m, and their dependence on the photon
energy. }\label{c1}
\end{figure}
\begin{figure}[h]
\centerline{\epsfig{file=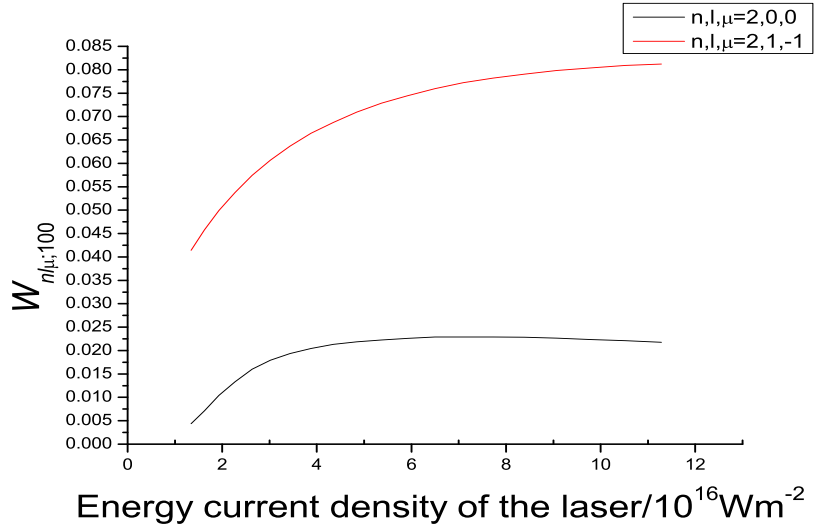,width=12cm, height=8cm}}
\caption{Transition probabilities of a hydrogen atom interacting
with a circularly polarized laser of $\hbar\omega=0.296$eV, and
their dependence on the laser intensity. }\label{c2}
\end{figure}

Fig.\ref{c1} shows that the spectrum is continuous. No discrete
sharp resonance peaks appear. If one fits the spectrum by
(\ref{3}), $\eta$ may take any real number in a wide range, not
necessarily be an integer. The transition is non-integer. While
fig.\ref{c2} shows that the transition probability is not
proportional to an integer power of the laser intensity. It means,
the interaction between the laser and the atom cannot be reduced
to the interaction of individual photons with the atom separately.
The interaction is between the atom and the laser as a whole. This
scenery is quite different from the regularity one saw in the weak
light (including weak laser) spectroscopy, therefore has to be
checked by new experiments. One may observe the radiation of the
atom when it is irradiated by an intense laser. This is the laser
Raman effect. In this way, the changes of distributions of atoms
among various energy levels, and therefore their transition
probabilities, are measured. Although there is not any separate
resonance, fig.\ref{c1} still shows complex structure in the
spectrum. It is interesting to find out the information exposed by
this kind of structure.

\section{ Laser photo-ionizations  \label{s3}}

The photo-ionization or the photoelectric effect is the transition
of the electron from the ground state to the ionized state, when
it is irradiated by light. The photo-ionization by an intense
circularly polarized laser may be handled by the method proposed
in \cite{z}. Some preliminary results obtained by this method have
been reported in \cite{z}-\cite{l}. Here we would analyze it from
the view point of non-integer quantum transition.

It is shown in \cite{z}, that the energy of the ionized electron
(photo-electron) is
\begin{eqnarray}
E_{f0}=E_i-\mu\hbar\omega\ ,\label{31}
\end{eqnarray}
and the transition probability per unit time is
\begin{eqnarray}
P=\frac{2\pi}{\hbar}\left|H_{f0,i}^{\prime\prime}\right|^2\rho\
,\label{32}
\end{eqnarray}
with
\begin{eqnarray}
H_{f0,i}^{\prime\prime}\equiv\int\psi_{f0}({\textit{\textbf
r}})\hat{H}^{\prime\prime}\phi_i({\textit{\textbf r}}){\rm
d}{\textit{\textbf r}}\ .\label{33}
\end{eqnarray}
$\phi_i({\textit{\textbf r}})$ in (\ref{33}) is an eigenfunction
of $\hat{H}_{\rm ps}$, satisfying (\ref{21}). $E_i$ in (\ref{31})
is the corresponding eigenvalue. $\psi_{f0}({\textit{\textbf r}})$
is the eigenfunction of $\hat{H}_0^\prime$, with eigenvalue
$E_{f0}+\mu\hbar\omega$, therefore is a projection of the Coulomb
wave function onto the subspace with definite magnetic quantum
number $\mu$, and describes the ionized electron.

In the weak light limit, $A\rightarrow 0$, $\phi_i$ approaches an
eigenfunction of $\hat{H}_0^\prime$, which is also the ground
state eigenfunction of $\hat{H}_0$ with zero magnetic quantum
number; and $E_i$ approaches the corresponding eigenvalue. They
are independent of $A$. For the hydrogen atom, they are
$\psi_{100}$ and $-b$ respectively, $b$ is the binding energy of
the electron in the ground state hydrogen atom. In the case of
$ka_0<< 1$, we have (\ref{280}), therefore the selection rule
(\ref{281}) works. These limits make the energy (\ref{31}) of the
photoelectron be
\begin{eqnarray}
E_{f0}=\hbar\omega-b\ , \label{34}
\end{eqnarray}
and the transition probability $P$ proportional to the light
intensity. This example shows, in the weak light limit, the
photo-ionization has the following distinct characters:

\vskip 0.1in \noindent C1.There is a critical frequency for a
given system.The light with frequency lower than this critical
value cannot eject any electron from the system.

\noindent C2.The light with frequency higher than this critical
value can ionize the system, the energy of the ejected electron
increases linearly with the increasing of the frequency but is
independent of the intensity of the light.

\noindent C3.The intensity of the photo-electric current is
proportional to the intensity of the light.

\vskip 0.1in \noindent This is exactly the experimental knowledge
on photo-ionization, people had before the discovery of the laser.
Based on this knowledge and guided by his idea of light quanta,
one hundred years ago, Einstein \cite{e} found his famous formula
(\ref{34}) and the idea that the light-atom interaction may be
reduced to the interactions between photons and atoms. In this
way, he explained the above experimental characters of
photo-ionization. This was a crucial step towards the discovery of
quantum mechanics. Now we see, all of these experimental
characters, as well as the Einstein formula (\ref{34}), together
with his idea that photons interact with atoms independently, are
the perturbation results of quantum mechanics in the weak light
limit. What will be the scenery when the light becomes an intense
laser?

\begin{figure}[h]
\centerline{\epsfig{file=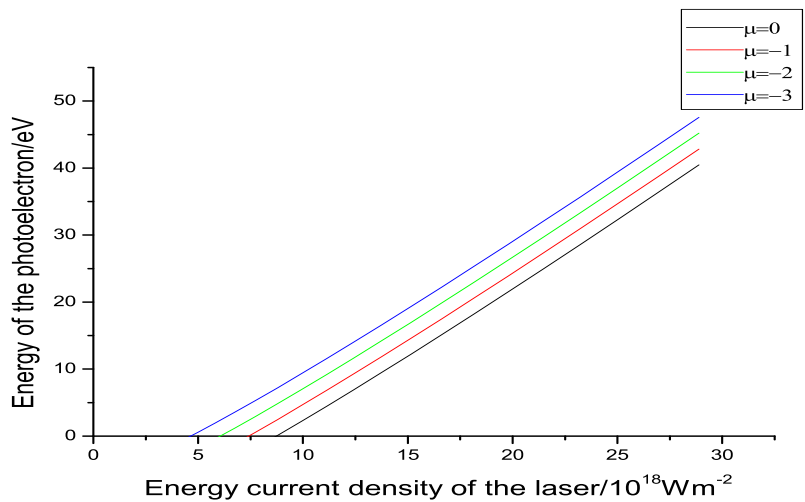,width=12cm, height=7.5cm}}
\caption{The energy of a photoelectron ejected from the ground
state hydrogen atom by a circularly polarized laser of
$\hbar\omega=2.37$eV, and its dependence on the laser intensity.
}\label{c3}
\end{figure}

If one puts $E_1=-b$ , and $E_2=E_{f0}$, the formula (\ref{34})
becomes (\ref{1}) with the positive sign on the right. Therefore,
the Einstein formula is a predecessor of the Bohr condition. Soon
after the discovery of laser, people observed the MPI\cite{v} and
the ATI\cite{a}. Einstein formula was generalized to be
\begin{eqnarray}
E_{f0}=N\hbar\omega-b\ , \label{35}
\end{eqnarray}
with $N$ being an positive integer. This is something like a
special case of the generalized Bohr condition (\ref{2}) and may
be deduced by the higher order perturbation. (\ref{31}) would be
an exact expression of the photoelectron energy, if $E_i$ in it is
solved from (\ref{21}) exactly. Defining
\begin{eqnarray}
\eta\equiv\frac{E_i+b}{\hbar\omega}-\mu\ ,
\end{eqnarray}
one may write (\ref{31}) in the form
\begin{eqnarray}
E_{f0}=\eta\hbar\omega-b\ .\label{36}
\end{eqnarray}
It is a special case of (\ref{3}). Here we see, $\eta$ is an
integer only when $E_i+b-\mu\hbar\omega$ is an integer multiple of
the photon energy $\hbar\omega$. This will not necessarily be the
case for a non-perturbation interaction between the atom and an
intense laser. It means the photo-ionization will be a non-integer
quantum transition. This is the true non-perturbation effect.
Using the numerical solution of (\ref{21}) obtained in the last
section, we find the light intensity dependence of the
photoelectron energy. The numerical result is shown in
fig.\ref{c3}.

\begin{figure}[h]
\centerline{\epsfig{file=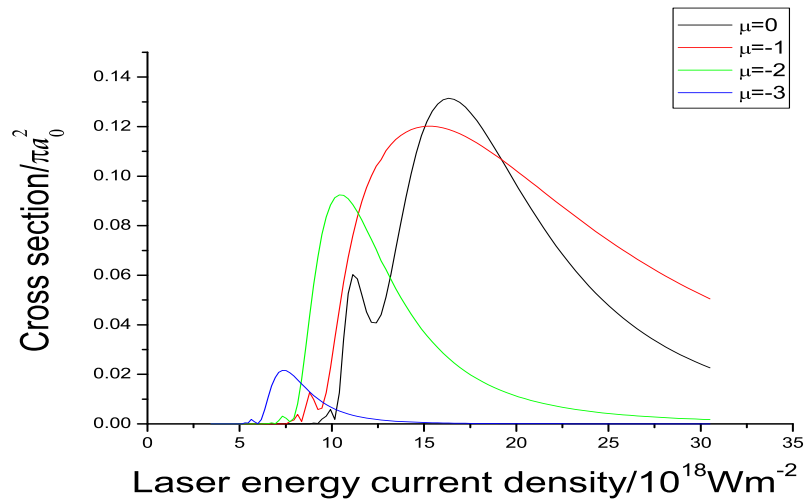,width=12cm, height=7.5cm}}
\caption{Cross section of the photo-ionization of a ground state
hydrogen atom irradiated by a circularly polarized laser of
$\hbar\omega=2.37$eV, and its dependence on laser intensity.
}\label{c4}
\end{figure}

The transition probability (\ref{32}) may be expressed in the form
of cross section. It is the formula (14) or (15) in \cite{z}.
Applying it to the photo-ionization of the hydrogen atom
irradiated by a circularly polarized laser, under the condition
$ka_0<<1$, we obtain the cross section
\begin{eqnarray}
\sigma=\frac{16\alpha
v}{ka_0c}\sum_{l=|\mu|}\left|\beta_l(i)\right|^2\label{37}
\end{eqnarray}
in unit of $\pi a_0^2$. Here, $v=\sqrt{2E_{f0}/m}$ is the velocity
of the photoelectron, and $\beta_l(i)$ is an elementary but some
what lengthy and tedious expression, containing integrals of the
type
\begin{eqnarray}
\int_0^\infty {\rm e}^{-st}t^{u-1}{\rm F}(a_1,c_1,t){\rm
F}(a_2,c_2,qt){\rm d}t=\Gamma(u)s^{-u}{\rm
F}_2(u,a_1,a_2,c_1,c_2,s^{-1},\frac{q}{s}) \ .
\end{eqnarray}
The integral has been analytically worked out. There are two
confluent hypergeometric functions F on the left. One is from the
radial wave function of the electron in the hydrogen atom, and
another is from the Coulomb wave function of the outgoing
electron. ${\rm F}_2$ on the right is the Appell's hypergeometric
function of the second class  in two variables\cite{er}. In our
problem here, it degenerates into a polynomial in two variables.
Therefore the calculation of $\beta_l(i)$ becomes finite, if the
expansion (\ref{23}) is truncated. The calculated cross section
and its dependence on the laser intensity is shown in
fig.\ref{c4}.

We see from fig.\ref{c3}, the energy of the photo-electron
increases with the increasing of the light intensity. The critical
frequency is not absolute. Even though  the frequency of the
incident light is lower than the critical frequency, the electron
may still be ejected, if the light is intense enough. The
characters C2 and C1, together with the formula (\ref{34}), are
not true for laser photo-ionization. Furthermore, the formula
(\ref{35}) is not true either, if the incident laser is very
strong. In this case, it has to be substituted by (\ref{36}), with
non-integer $\eta$. The transition in photo-ionization becomes
non-integer. However, one may still see an apparent quantum
character in fig.\ref{c3}. That is, the energy difference between
photo-electrons with different magnetic quantum number $\mu$ is an
integer multiple of the quantum energy $\hbar\omega$. From
fig.\ref{c4} we see, the cross sections depend on the light
intensity nonlinearly. It means that the character C3 is not true
for laser photo-ionization. The interaction between light and
atoms cannot be reduced to the independent interactions between
photons and atoms. Atoms interact with the laser as a whole.

\section{Omitted effects \label{s4}}
We omitted some effects in the foregoing sections. Here let us say
a few words on them.
\subsection{The motion of the nucleus}
The hydrogen atom consists of a proton and an electron. To
consider the motion of the proton, one has to change (\ref{13})
into
\begin{eqnarray}
\hat{H}_0=\frac{\hat{p}_1^2}{2m_1}+\frac{\hat{p}_2^2}{2m_2}+V(r)\
,\label{130}
\end{eqnarray}
and (\ref{14}) into
\begin{eqnarray}
\hat{H}^\prime &=&\frac{eA}{m_1} [\hat{p}_{1x}\cos(kz_1-\omega
t)+\hat{p}_{1y}\sin(kz_1-\omega t)]+\frac{e^2A^2}{2m_1} \nonumber
\\&-&\frac{eA}{m_2}[\hat{p}_{2x}\cos(kz_2-\omega
t)+\hat{p}_{2y}\sin(kz_2-\omega t)]+\frac{e^2A^2}{2m_2} \
.\label{140}
\end{eqnarray}
Subscripts 1 and 2 denote the electron and the proton
respectively. Substituting them into (\ref{12}) and (\ref{15}),
and performing the transformation
\begin{eqnarray}
\Psi({\textit{\textbf r}}_1,{\textit{\textbf r}}_2,t)={\rm
e}^{{\rm i}\omega(\hat{L}_{1z}+\hat{L}_{2z})t
/\hbar}\Phi({\textit{\textbf r}}_1{\textit{\textbf r}}_2,t)\ ,
\label{160}
\end{eqnarray}
one obtains again the time independent pseudo-Schr\"odinger
equation (\ref{17}). But now one has to substitute
\begin{eqnarray}
\hat{H}_0^\prime=\hat{H}_0+\omega\sum_{j=1}^2\hat{L}_{jz}\label{190}
\end{eqnarray}
and
\begin{eqnarray}
\hat{H}^{\prime\prime}&=&\frac{eA}{m_1}
[\hat{p}_{1x}\cos(kz_1)+\hat{p}_{1y}\sin(kz_1)]+\frac{e^2A^2}{2m_1}
\nonumber
\\&-&\frac{eA}{m_2}[\hat{p}_{2x}\cos(kz_2)+\hat{p}_{2y}\sin(kz_2)]+\frac{e^2A^2}{2m_2}
\label{200}
\end{eqnarray}
into the pseudo-Hamiltonian (\ref{18}). A further transformation
\begin{eqnarray}
\Phi=\exp\left[-{\rm
i}k\frac{(m_1z_1+m_2z_2)(\hat{L}_{1z}+\hat{L}_{2z})}{(m_1+m_2)\hbar}\right]\Phi_{\rm
e}
\end{eqnarray}
brings (\ref{17}) to
\begin{eqnarray}
{\rm i}\hbar\frac{\partial\Phi_{\rm e}}{\partial t}=\hat{H}_{\rm
e}\Phi_{\rm e}\ , \label{170}
\end{eqnarray}
with the effective Hamiltonian
\begin{eqnarray}
\hat{H}_{\rm
e}&=&\sum_{j=1}^2\left[\frac{\hat{p}_{jx}^2+\hat{p}_{jy}^2+(\hat{p}_{jz}
-m_jk\frac{\hat{L}_{1z}+\hat{L}_{2z}}{m_1+m_2})^2}{2m_j}
+\omega\hat{L}_{jz}\right]+V(r)\nonumber\\
&+&\frac{eA}{m_1}
\left[\hat{p}_{1x}\cos\left(km_2\frac{z_1-z_2}{m_1+m_2}\right)
+\hat{p}_{1y}\sin\left(km_2\frac{z_1-z_2}{m_1+m_2}\right)\right]+\frac{e^2A^2}{2m_1}
\nonumber
\\&-&\frac{eA}{m_2}\left[\hat{p}_{2x}\cos\left(km_1\frac{z_2-z_1}{m_1+m_2}\right)
+\hat{p}_{2y}\sin\left(km_1\frac{z_2-z_1}{m_1+m_2}\right)\right]+\frac{e^2A^2}{2m_2}\
.\label{180}
\end{eqnarray}
Introducing the center of mass coordinates ${\textit{\textbf
R}}\equiv(m_1{\textit{\textbf r}}_1+m_2{\textit{\textbf r}}_2)/M$
and the relative coordinates ${\textit{\textbf
r}}\equiv{\textit{\textbf r}}_1-{\textit{\textbf r}}_2$,
 we have the total momentum
$\hat{\textit{\textbf P}}=\hat{\textit{\textbf
p}}_1+\hat{\textit{\textbf p}}_2$, the relative momentum
$\hat{\textit{\textbf p}}=m(\hat{\textit{\textbf
p}}_1/m_1-\hat{\textit{\textbf p}}_2/m_2)$, the angular momentum
$\hat{\textit{\textbf L}}_c={\textit{\textbf
R}}\times\hat{\textit{\textbf P}}$ of the center of mass, and the
angular momentum $\hat{\textit{\textbf L}}={\textit{\textbf
r}}\times\hat{\textit{\textbf p}}$ around the center of mass.
$M=m_1+m_2$ is the total mass, and $m=m_1m_2/(m_1+m_2)$ is the
reduced mass. In these coordinates, the effective Hamiltonian
(\ref{180}) has the form
\begin{eqnarray}
\hat{H}_{\rm e}&=&\frac{\hat{\textit{\textbf
P}}_x^2+\hat{\textit{\textbf P}}_y^2+[\hat{\textit{\textbf P}}_z
-k(\hat{L}_{cz}+\hat{L}_z)]^2}{2M} +\omega\hat{L}_{cz}\nonumber\\
&+&\frac{p^2}{2m}+V(r)+\omega\hat{L}_z+\frac{eA}{m}\hat{p}_x
+\frac{e^2A^2}{2m}\ ,\label{201}
\end{eqnarray}
at the limit of $ka_0\rightarrow0 $. The sum of the last five
terms relates to the relative motion only, and equals the
pseudo-Hamiltonian (\ref{18}) (together with (\ref{20})) at the
same limit of $ka_0\rightarrow0$, if $m$ there is also understood
to be the reduced mass instead of the electron mass. The first two
terms mainly relate to the motion of the center of mass. Only
$\hat{L}_z$ in the first term relates to the relative motion. But
the big mass $M$ on the denominator makes its contribution be much
less than that of the last five terms. Therefore, one needs only
to consider the sum of the last five terms in (\ref{201}), for the
problem of relative motion between the electron and the proton in
hydrogen atom, irradiated by a circularly polarized laser. The
correction of the nucleus motion is again the substitution of the
reduced mass for the electron mass. It is tiny. The first two
terms in (\ref{201}) govern the motion of the hydrogen atom as a
whole. They have to be considered if one is interested in the
motion of ionized electrons, for example, their angular
distributions.
\subsection{The quantization of the electromagnetic field}
In a complete theory, the electromagnetic field has to be
quantized. In the Coulomb gauge, it is to let the vector potential
be an operator $\hat{\textit{\textbf{A}}}$ and define commutators
between its components. Introducing a complete set of vector
functions $[{\textit{\textbf{A}}}_\iota({\textit{\textbf{r}}})]$,
satisfying the Helmholtz equations
\begin{eqnarray}
\nabla^2{\textit{\textbf{A}}}_\iota({\textit{\textbf{r}}})
+\frac{\omega_\iota^2}{c^2}{\textit{\textbf{A}}}_\iota({\textit{\textbf{r}}})=0
\end{eqnarray}
and the orthonomal conditions
\begin{eqnarray}
\int{\textit{\textbf{A}}}_\iota^*({\textit{\textbf{r}}})\cdot
{\textit{\textbf{A}}}_\iota^\prime({\textit{\textbf{r}}}){\rm
d}{\textit{\textbf{r}}} = \delta_{\iota\iota^\prime}\ ,
\end{eqnarray}
one may expand the self-adjoint operator
\begin{eqnarray}
\hat{\textit{\textbf{A}}}({\textit{\textbf{r}}},t)
=\sum_\iota\sqrt{\frac{\hbar}{2\epsilon_0\omega_\iota}}
\left[\hat{b}_\iota{\textit{\textbf{A}}}_\iota({\textit{\textbf{r}}})
+\hat{b}_\iota^\dag{\textit{\textbf{A}}}_\iota^*({\textit{\textbf{r}}})
\right]\ ,
\end{eqnarray}
$\epsilon_0$ is the dielectric constant for the vacuum. The
quantization condition is the commutators
\begin{eqnarray}
\hat{b}_\iota\hat{b}_{\iota^\prime}-\hat{b}_{\iota^\prime}\hat{b}_\iota
=\hat{b}_\iota^\dag\hat{b}_{\iota^\prime}^\dag-\hat{b}_{\iota^\prime}^\dag
\hat{b}_\iota^\dag=0\ ,\;\;\;\;\mbox{and}\;\;\;\;
\hat{b}_\iota\hat{b}_{\iota^\prime}^\dag-\hat{b}_{\iota^\prime}^\dag
\hat{b}_\iota=\delta_{\iota\iota^\prime}\ .
\end{eqnarray}
The vacuum state $|0\rangle$ is defined by
\begin{eqnarray}
\hat{b}_\iota|0\rangle=0\;\;\;\;\;\;\mbox{for all $\iota$}\ .
\end{eqnarray}
This is the quantization around the vacuum. Classically, the
vacuum is described by a vector potential
${\textit{\textbf{A}}}_0({\textit{\textbf{r}}},t)=0$, which is a
trivial solution of the D'Alembert equation. It suggests, that one
may also quantize the theory around another classical solution
${\textit{\textbf{A}}}_c({\textit{\textbf{r}}},t)$, for example
the solution (\ref{11}), of the D'Alembert equation.  Defining
$\hat{\textit{\textbf{A}}}^\prime({\textit{\textbf{r}}},t)
=\hat{\textit{\textbf{A}}}({\textit{\textbf{r}}},t)
-{\textit{\textbf{A}}}_c({\textit{\textbf{r}}},t)$, expanding
\begin{eqnarray}
{\textit{\textbf{A}}}_c({\textit{\textbf{r}}},t)
=\sum_\iota\sqrt{\frac{\hbar}{2\epsilon_0\omega_\iota}}
\left[c_\iota{\textit{\textbf{A}}}_\iota({\textit{\textbf{r}}})
+c_\iota^*{\textit{\textbf{A}}}_\iota^*({\textit{\textbf{r}}})
\right]\ ,
\end{eqnarray}
we have
\begin{eqnarray}
\hat{\textit{\textbf{A}}}^\prime({\textit{\textbf{r}}},t)
=\sum_\iota\sqrt{\frac{\hbar}{2\epsilon_0\omega_\iota}}
\left[\hat{b}_\iota^\prime{\textit{\textbf{A}}}_\iota({\textit{\textbf{r}}})
+\hat{b}_\iota^{\prime\dag}{\textit{\textbf{A}}}_\iota^*({\textit{\textbf{r}}})
\right]\ ,
\end{eqnarray}
with
\begin{eqnarray}
\hat{b}^\prime_\iota=\hat{b}_\iota-c_\iota\ .
\end{eqnarray}
Since $c_\iota$ are c-numbers, $\hat{b}_\iota^\prime$ and
$\hat{b}_\iota^{\prime\dag}$ have the same commutators as those
for $\hat{b}_\iota$ and $\hat{b}_\iota^\dag$. They are
\begin{eqnarray}
\hat{b}_\iota^\prime\hat{b}_{\iota^\prime}^\prime-\hat{b}_{\iota^\prime}^\prime\hat{b}_\iota^\prime
=\hat{b}_\iota^{\prime\dag}\hat{b}_{\iota^\prime}^{\prime\dag}-\hat{b}_{\iota^\prime}^{\prime\dag}
\hat{b}_\iota^{\prime\dag}=0\ ,\;\;\;\;\mbox{and}\;\;\;\;
\hat{b}_\iota^\prime\hat{b}_{\iota^\prime}^{\prime\dag}-\hat{b}_{\iota^\prime}^{\prime\dag}
\hat{b}_\iota^\prime=\delta_{\iota\iota^\prime}\ .
\end{eqnarray}
The quantization condition for the electromagnetic field
$\hat{\textit{\textbf{A}}}^\prime$ around a classical field
${\textit{\textbf{A}}}_c$ is therefore the same as that for the
field $\hat{\textit{\textbf{A}}}$ around the classical vacuum
${\textit{\textbf{A}}}_0=0$. However, the 'vacuum' state is now
changed to $|c\rangle$, satisfying
$\hat{b}_\iota^\prime|c\rangle=0$. This is
\begin{eqnarray}
\hat{b}_\iota|c\rangle=c_\iota|c\rangle\ ,
\end{eqnarray}
showing that $|c\rangle$ is a coherent state with non-zero
amplitude(s) $c_\iota$. In the classical limit it is
${\textit{\textbf{A}}}_c$ itself.

The interaction operator between the electromagnetic field and a
non-relativistic electron is
\begin{eqnarray}
\frac{e}{m}\hat{\textit{\textbf{A}}}\cdot\hat{\textit{\textbf{p}}}
+\frac{e^2\hat{\textit{\textbf{A}}}^2}{2m}=\frac{e}{m}{\textit{\textbf{A}}}_c
\cdot\hat{\textit{\textbf{p}}}+\frac{e^2A_c^2}{2m}+\frac{e}{m}
\hat{\textit{\textbf{A}}}^\prime\cdot\hat{\textit{\textbf{p}}}
+\frac{e^2\hat{\textit{\textbf{A}}}^{\prime2}}{2m}+\frac{e^2
\hat{\textit{\textbf{A}}}^\prime\cdot{\textit{\textbf{A}}}_c}{m}\
.
\end{eqnarray}
If ${\textit{\textbf{A}}}_c$ represents an intense laser, the
first two terms on the right would be large, its effect has to be
treated non-perturbatively. It is the main part of the problem.
This is what we have done in the above sections. For a few
fluctuations of the electromagnetic field around the laser, the
remaining terms on the right of this equation are small, and may
be considered by perturbation, if it is needed.
\subsection{The relativity and spin effects of the electron}
The method used above may be applied to a relativistic particle
system with spin as well. The way is to use the total angular
momentum $\hat{\textit{\textbf{J}}}$, instead of the orbital
angular momentum $\hat{\textit{\textbf{L}}}$, in the
transformation (\ref{16}). In this way, the transformation reads
\begin{eqnarray}
\Psi={\rm e}^{{\rm i}\omega\hat{J}_zt /\hbar}\Phi\ .\label{260}
\end{eqnarray}
The relativity is not important in most problems. One can easily
consider the electron spin by applying this transformation in
solving the Pauli equation for electrons in an atom, irradiated by
the circularly polarized laser, whenever he is interested in the
problem of electron polarization in the process.
\section{Conclusion}
We see, a laser may not only induce MPI and ATI, but also cause
non-integer transitions, if it is strong enough. The later can
only be handled by non-perturbation method, and therefore is a
non-perturbation effect.

One hundred years ago, people knew very few about the
photo-electric effects. There was not a laser. People could see
the effect only when irradiating matter by usual light. It is very
weak from the present view point. But, just under this condition,
the distinct characters C1-C3 shown in section \ref{s3} appear. It
was these distinct characters made Einstein find the light quanta
by his keen insight, which was one of the important steps towards
the discovery of quantum mechanics. Several decades later, people
predicted and constructed the laser by the guide of quantum
mechanics. Now we see, again by the guide of quantum mechanics, if
one irradiates matter by intense laser, very fruitful and complex
phenomena will appear, and those distinct characters disappear. We
are fortunate, that people discovered usual light instead of the
laser first, so that Einstein could see the distinct characters
and discover the light quanta one hundred years ago. We learn from
this history, that sometimes simple experimental phenomena may
expose essentials; on the contrary, too fruitful experimental data
may conceal essential points. In any case, a keen insight is
always important.

\end{document}